\documentclass[
]{ceurart}

\usepackage{listings}
\usepackage{xurl}

\usepackage{breakurl}

\begin{document}



\title{OSINT – or BULLSHINT? Exploring Open-Source Intelligence tweets about the Russo-Ukrainian War}

\author[1]{Johannes Niu}[%
orcid=,
email=,
url=,
]
\address[1]{Independent Researcher}

\author[2]{Mila Stillman}[%
orcid=0000-0001-7116-9338,
email=mila.stillman@hm.edu,
url=,
]
\address[2]{Munich University of Applied Sciences,
Lothstr. 64,
80335 Munich, Germany}

\author[2]{Anna Kruspe}[%
orcid=0000-0002-2041-9453,
email=anna.kruspe@hm.edu,
url=http://cs.hm.edu/~kruspe,
]


\begin{abstract}
  This paper examines the role of Open Source Intelligence (OSINT) on Twitter regarding the Russo-Ukrainian war, distinguishing between genuine OSINT and deceptive misinformation efforts, termed ``BULLSHINT.'' Utilizing a dataset spanning from January 2022 to July 2023, we analyze nearly 2 million tweets from approximately 1,040 users involved in discussing real-time military engagements, strategic analyses, and misinformation related to the conflict. Using sentiment analysis, partisanship detection, misinformation identification, and Named Entity Recognition (NER), we uncover communicative patterns and dissemination strategies within the OSINT community. Significant findings reveal a predominant negative sentiment influenced by war events, a nuanced distribution of pro-Ukrainian and pro-Russian partisanship, and the potential strategic manipulation of information. Additionally, we apply community detection techniques, which are able to identify distinct clusters partisanship, topics, and misinformation, highlighting the complex dynamics of information spread on social media. This research contributes to the understanding of digital warfare and misinformation dynamics, offering insights into the operationalization of OSINT in geopolitical conflicts. 
\end{abstract}


\maketitle

\section{Introduction}
Since Russia's full-scale invasion of Ukraine on February 24, 2022, following years of simmering conflict that began in 2014, social media platforms have been battlegrounds for narrative and interpretive authority. Platforms like Twitter have played crucial roles, both in the Ukrainian Maidan movement of 2013/2014 and in the ongoing Russian disinformation campaigns. These platforms have been instrumental in influencing public opinion and military strategies through the dissemination of both official information and targeted misinformation campaigns.

The invasion has amplified the role of social media in modern warfare, marking this conflict as the "first social media war," where online narratives can significantly impact the geopolitical landscape. Open Source Intelligence (OSINT) communities have emerged as pivotal players in this digital arena, utilizing publicly available data to analyze, fact-check, and counter misinformation. High-profile cases like the MH17 incident have highlighted the effectiveness of OSINT in debunking false narratives propagated by state actors.

This study focuses on the dynamics of OSINT communities on Twitter, aiming to discern genuine OSINT from deliberate misinformation, termed ``BULLSHINT.'' By examining the behavior and network interactions of these users, the research seeks to understand how misinformation spreads and how it can be effectively countered, thereby contributing to a more informed global response to the challenges posed by hybrid warfare and the digital manipulation of public discourse.

The paper is structured as follows: 

\section{Related work}
Recent research on social media's role in the Russo-Ukrainian war has largely focused on sentiment analysis, public opinion, and the detection of misinformation in OSINT-related content. \cite{agarwal_exploring_2022, poleksicSentimentTweetsRussoUkrainian2023, ingoleContentBasedStudyTweet2023} utilize sentiment analysis tools like VADER to assess public emotions and opinions expressed on Twitter, revealing a predominance of negative sentiments and the evolution of public discourse surrounding the conflict.

Further investigations into misinformation and narrative manipulation are highlighted in \cite{hanleyHappenstanceUtilizingSemantic2023, hanleySpecialOperationQuantitative2023, parkChallengesOpportunitiesInformation2022, Geissler2023RussianPropaganda}. These studies examine how different media ecosystems and social media bots influence public opinion and disseminate pro-Russian propaganda, using methods ranging from network analysis to topic extraction.

The role of OSINT in countering misinformation has become increasingly significant, as demonstrated by the analysis of user-generated content and the verification efforts by entities like Bellingcat. \cite{Kemp2022OSINT} and \cite{Hayman2023OpenSource} emphasize the strategic implications of OSINT in modern warfare, highlighting its effectiveness in real-time information dissemination and its impact on military and public decision-making.

This study builds on these findings by analyzing Twitter data to identify genuine OSINT activities versus ``BULLSHINT'' — misleading information presented as legitimate intelligence. By examining tweet interactions, text features, and community structures, this research aims to uncover patterns that help differentiate between authentic OSINT contributions and misinformation efforts. This approach not only contributes to understanding the dynamics of digital warfare but also assists in enhancing the reliability of OSINT practices in conflict zones.

\section{Data}
The dataset we analyze is the one presented in \cite{niu2024}. This dataset, assembled from Twitter and spanning from January 2022 to July 2023, focuses on Open Source Intelligence (OSINT) related to the Russo-Ukrainian war. Constructed using a two-step snowball sampling method, the dataset began with keyword searches for ``OSINT'' alongside ``Ukraine'' or ``Russia,'' and expanded through user interactions such as retweets and mentions to capture a broader community of OSINT contributors. This approach yielded nearly 2 million tweets from about 1,040 users, encompassing real-time military engagements, strategic analyses, and discussions around misinformation, which highlight the role of social media in shaping conflict narratives. The dataset’s rich multimedia content includes images, videos, and external links, offering a comprehensive view of the war’s coverage. For ongoing research, this dataset provides a critical resource for examining OSINT evolution, its impact on public perception, and misinformation dissemination during conflicts, supporting deeper analyses such as sentiment analysis and network behavior.

\section{Data analysis}
Based on this comprehensive dataset, we perform a range of analysis, focusing on two primary objectives:
\begin{enumerate}
    \item Revealing insights into the communication patterns of OSINT-associated users discussing the Russo-Ukrainian war.
    \item Creating informative aggregated user metrics that reflect the tweets of users via features retrieved from tweets.
\end{enumerate}
To achieve these objectives, we employ several analytical techniques:
\begin{itemize}
    \item Sentiment Analysis: Analyzing the sentiments expressed in tweets to understand the emotional tone and prevalent opinions within the OSINT community.
    \item Partisanship Detection: Investigating tweets for pro-Ukrainian or pro-Russian partisanship.
    \item Misinformation Detection: Analyzing tweets for the presence of misinformation to assess the reliability of the content being shared.
    \item Named Entity Recognition (NER): Extracting and classifying named entities mentioned in tweets to understand key subjects and entities being discussed.
\end{itemize}

To prepare the data for each analysis step, we perform the following preprocessing steps: Lowercasing; cleanup of URLs, user names, punctuation, special characters, and digits; stopword removal; tokenization; lemmatization and stemming. As some of these steps are language-specific, we focus on the subset of the data written in English, Ukrainian, Russian, Japanese, German, French, Italian, Spanish, and Portuguese, resulting in 1.6 million tweets. We then employ pre-trained classification models for each individual analysis.

\subsection{Sentiment}
Sentiments were analyzed via CardiffNLP's pretrained models: The one from \cite{cardifnlpSentiment} for English-language tweets, and the one from \cite{cardiffnlpSentimentMultiLingual} for other languages. The over-all results are shown in Figure \ref{}. We find that the majority of tweets has negative sentiment, which is to be expected for the war topic, but a large number also presents information in a factual, neutral manner. Around 9\% are positive, typically manifesting as tweets expressing hope or defiance (e.g. ``Slava Ukraini'').

\begin{figure}
    \centering
    \includegraphics[width=0.8\linewidth]{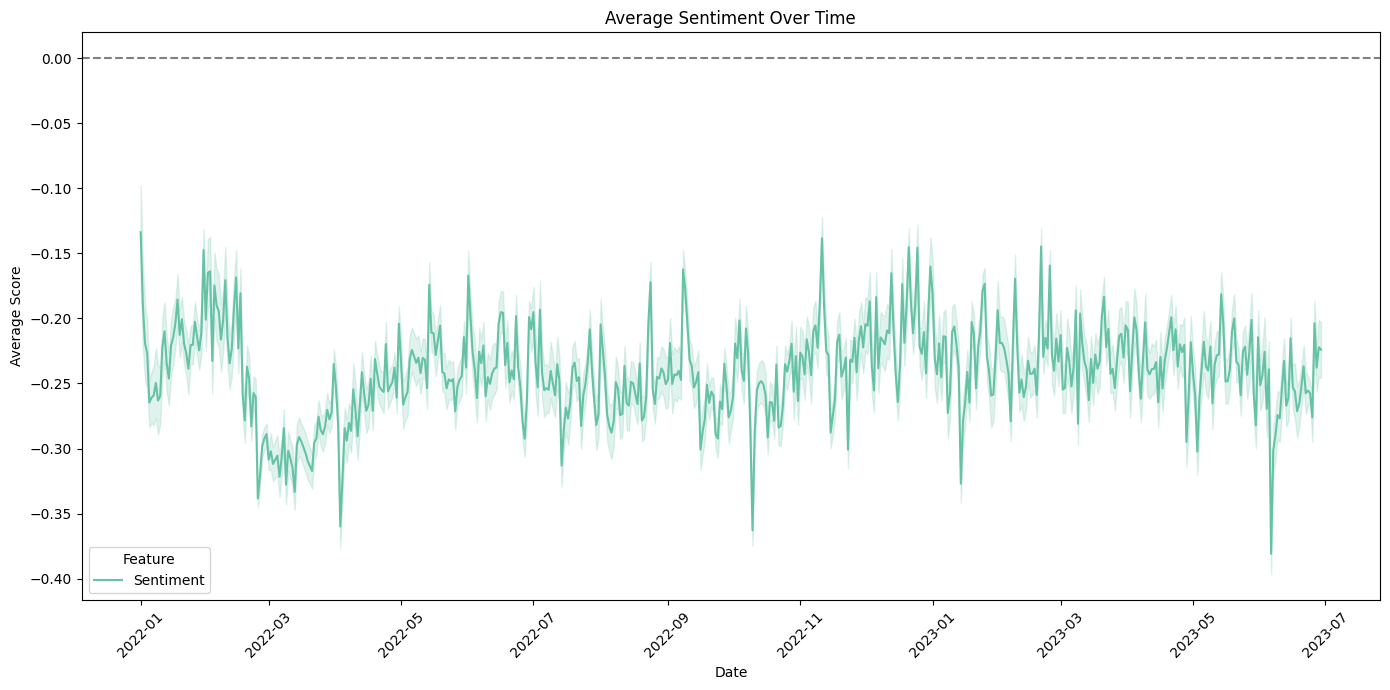}
    \caption{Development of sentiment values over time (0 - neutral, range between -1 and 1).}
    \label{fig:sentiment_over_time}
\end{figure}

Figure \ref{fig:sentiment_over_time} shows the development over time. Overall average sentiments are consistently negative, with a notable decline in February and March 2022, the beginning of the Russian incursion. Post-invasion, sentiment fluctuates but remains negative, with peaks around significant events, highlighting ongoing public concern and negativity. This suggests that the attack on Ukraine influenced users sentiment negatively, which hints at a general pro-Ukrainian sentiment in the population.

\begin{figure}
    \centering
    \includegraphics[width=0.8\linewidth]{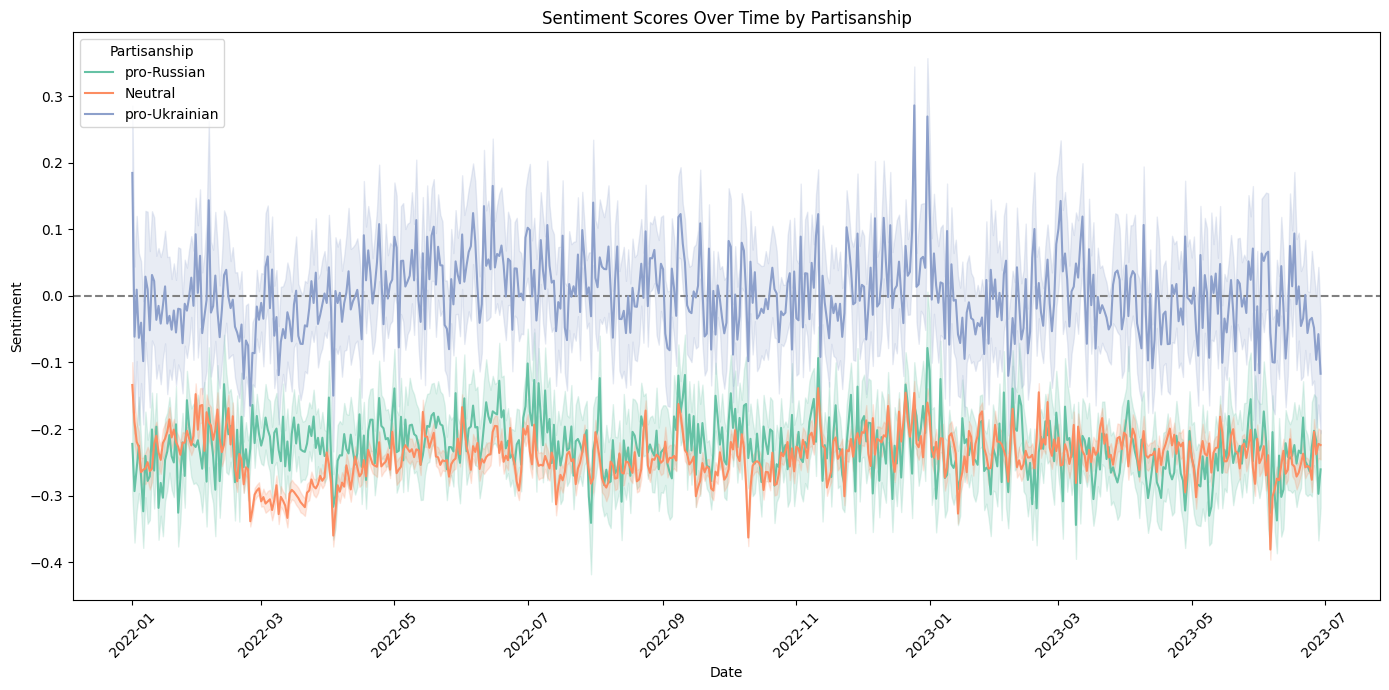}
    \caption{Development of sentiment over time within the partisanship groups.}
    \label{fig:sentiment_by_partisanship}
\end{figure}

\subsection{Partisanship}
To identify either a pro-Russian or pro-Ukrainian bias in tweet texts, a binary classification model from Huggingface was used\footnote{\url{https://huggingface.co/YaraKyrychenko/ukraine-war-pov}}. 
A neutral stance towards either country is detected in 81.7\% of tweets, reflecting factual content as opposed to expression of opinions in a majority of the OSINT space. Surprisingly, pro-Russian partisanship is slightly more frequent (8.4\%) than pro-Ukrainian opinions (5.9\%; 4.1\% are undefined), despite the ban of Twitter in Russia since February 2022. We do not observe major changes over time apart from a slight decline in pro-Russian content at the start of the war.

We separately analyze the sentiment in each partisanship class, shown in Figure \ref{fig:sentiment_by_partisanship}. It appears that pro-Ukrainian tweets generally have positive sentiment scores (reiterating the defiant stances described above), while neutral and pro-Russian tweets maintain negative sentiment scores. Neutral tweets show a drop towards more negative sentiments in early 2022, after which they stabilize at a negative level.


\subsection{Misinformation}
Detecting misinformation\footnote{Note: We use the term ``misinformation'' to encompass both misinformation, which is unintentionally incorrect information, and disinformation, which is incorrect information spread with the intent to deceive, as our models currently cannot distinguish the two.} without ground truth of what is correct and what is not is a difficult task. However, one of the most interesting questions in the field of OSINT lies in determining what information is accurate, and what messages are misinformation or propaganda masquerading as new insights. To obtain an estimate, we utilize two pre-trained HuggingFace models\footnote{\url{https://huggingface.co/spencer-gable-cook/COVID-19_Misinformation_Detector}}\footnote{\url{https://huggingface.co/FriedGil/distillBERT-misinformation-classifier}}, and average their softmax outputs (where 1 signifies ``real'' and 0 signifies ``fake''). This results in a fraction of 65.1\% above $0.66$, 25.7\% between $0.33$ and $0.66$, with 9.2\% below $0.33$ - i.e. fairly certain to be misinformation.

\begin{figure}
    \centering
    \includegraphics[width=0.8\linewidth]{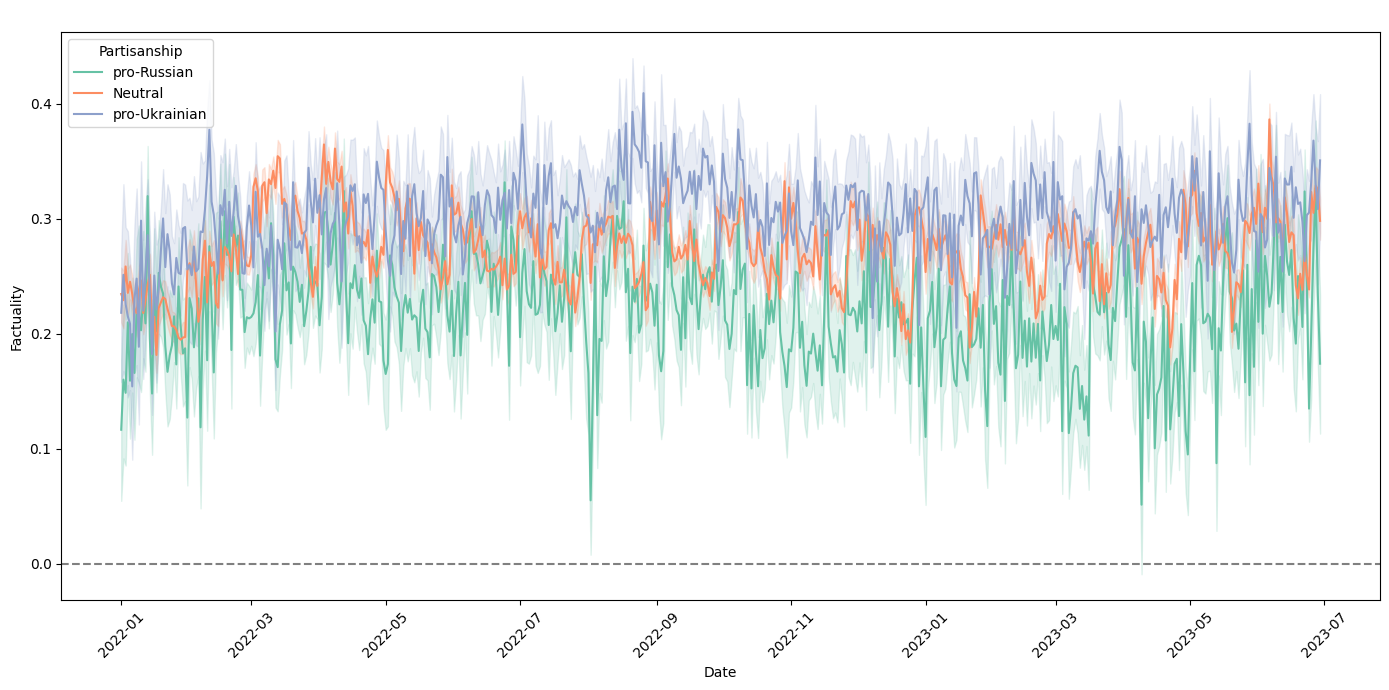}
    \caption{Detected development of factuality scores by partisanship over time (where 1 - real, 0 - fake).}
    \label{fig:misinfo_by_partisanship}
\end{figure}

An aggregation over time by partisanship classes, shown in Figure \ref{fig:misinfo_by_partisanship}, suggests that pro-Russian messages tend to distribute more misinformation than pro-Ukrainian ones, with neutral stances in the middle. We also see that misinformation in general fluctuates strongly over time, often reacting to war events. This is most pronounced for pro-Russian messages.



\subsection{Named entities}
Finally, we perform Named Entity Recognition on the data set using  babelscape \cite{tedeschi-etal-2021-wikineural-combined}. This is often helpful to obtain quick insights into the main topics of interest, e.g. in terms of persons, locations, and organizations. Almost 60\% of the tweet text contents contain named entities.

\begin{figure}
    \centering
    \includegraphics[width=0.8\linewidth]{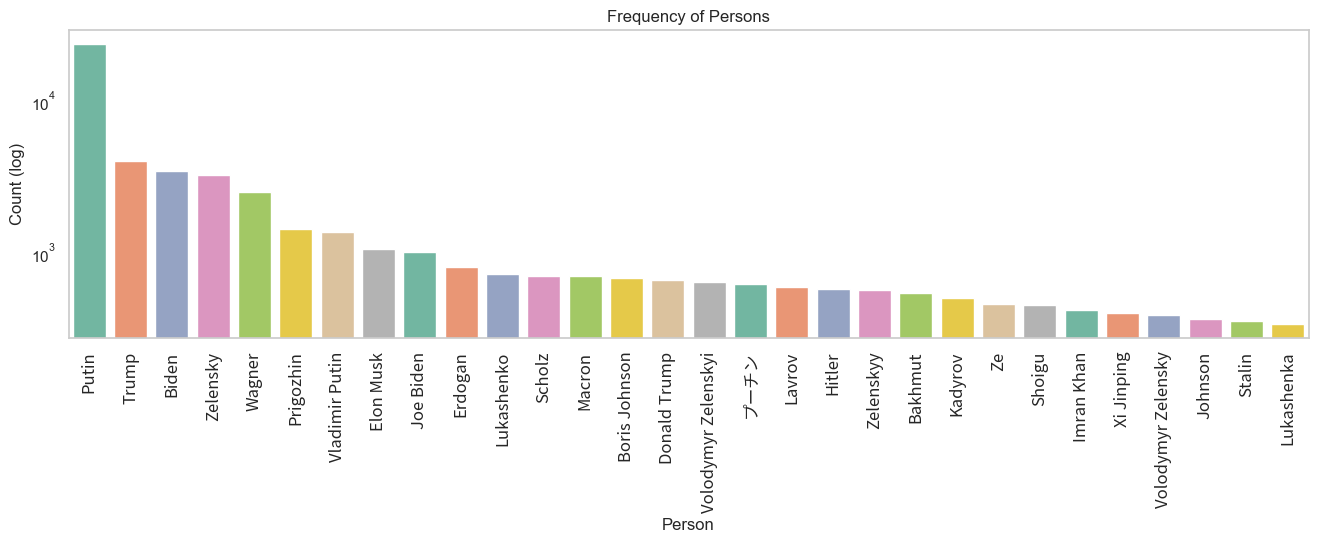}
    \caption{Most frequent named person entities.}
    \label{fig:ner_persons}
\end{figure}

Around 8.8\% of the collected tweets contain names of persons; the most frequent ones are shown in figure \ref{fig:ner_persons}. Most frequent among them, as expected, are political leaders and government officials of countries directly or indirectly involved in the war (``Putin'', ``Volodymyr Zelensky'', ``Lukashenko'', ``Lavrov'', ``Kadyrov''), and to a surprisingly high degree those of other significant world powers (``Trump'', ``Joe Biden'', ``Erdogan'', ``Xi Jinping'', ``Scholz'', ``Macron'', ``Boris Johnson'') as well as historical figures often used in comparisons (``Hitler'', ``Stalin''). We did not compress different spellings (e.g. ``Volodymyr Zelenskyi'' vs ``Volodymyr Zelensky''; ``Lukashenko'' vs ``Lukashenka''), different first and last name combinations (e.g. ``Trump'' vs ``Donald Trump''; ``Biden'' vs ``Joe Biden''), and nicknames (``Ze'' for Volodomyr Zelensky) as they may occur in different contexts and be used by different groups.

``Wagner'' is detected as a person’s name. In this context, however, the name refers to the private military complex of the same name.  ``Bakhmut'' is an erroneous recognition as well.


\begin{figure}
    \centering
    \includegraphics[width=0.8\linewidth]{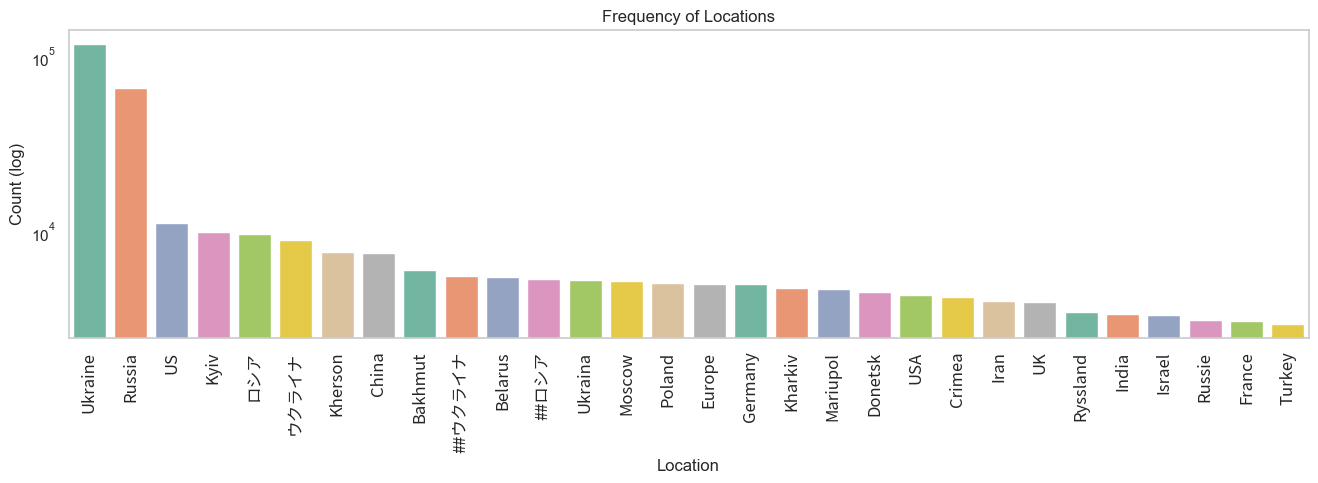}
    \caption{Most frequent named geographic entities.}
    \label{fig:ner_locations}
\end{figure}

\begin{figure}
    \centering
    \includegraphics[width=0.6\linewidth]{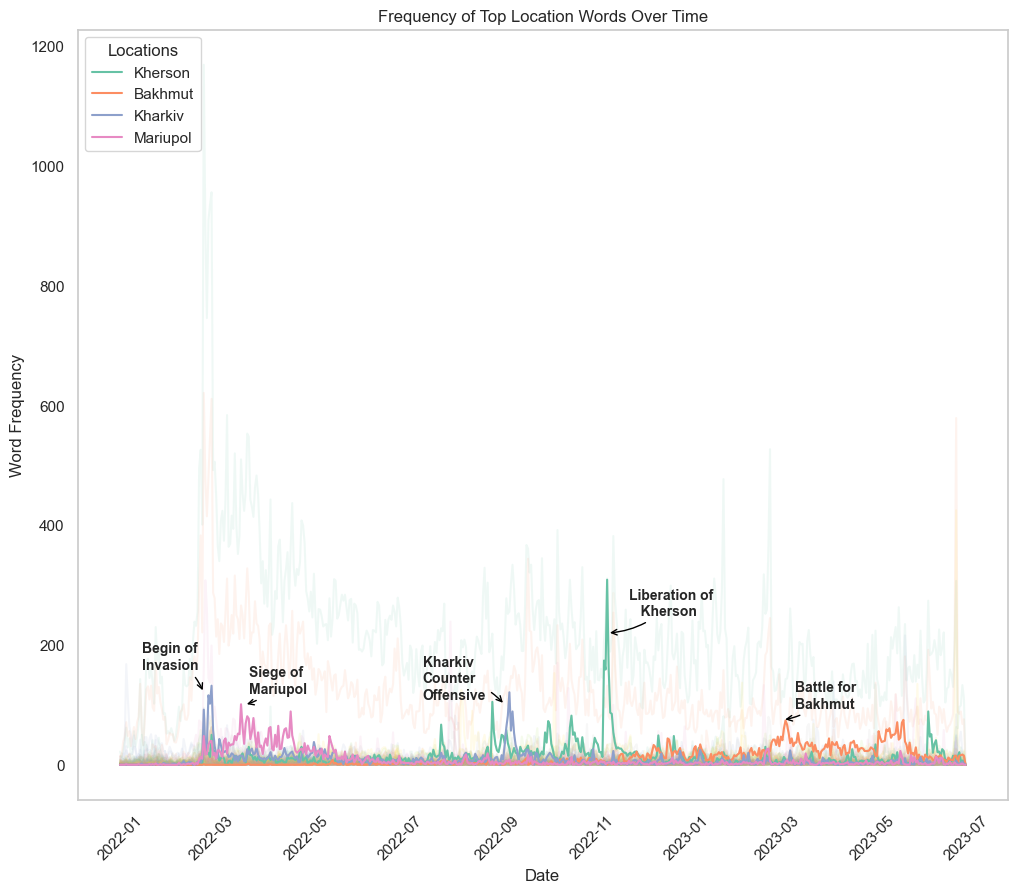}
    \caption{Development of select named geographic entities over time.}
    \label{fig:ner_locations_over_time}
\end{figure}

Geographic entity names are found in 21\% of the texts. Their distribution is shown in figure \ref{fig:ner_locations}. Country names dominate, with the first cities being ``Kyiv" in fourth in ``Kherson'' in 7th place (Bakhmut is recognized correctly here). Countries like ``China'', ``Poland'', or ``Germany'', as well as ``USA'', ``UK'' and ``Iran'' highlight the global influence of the conflict. Figure \ref{fig:ner_locations_over_time} shows the developments of city mentions over the course of the war, mainly corresponding to local events, but often already appearing earlier in the discussion.

\begin{figure}
    \centering
    \includegraphics[width=0.8\linewidth]{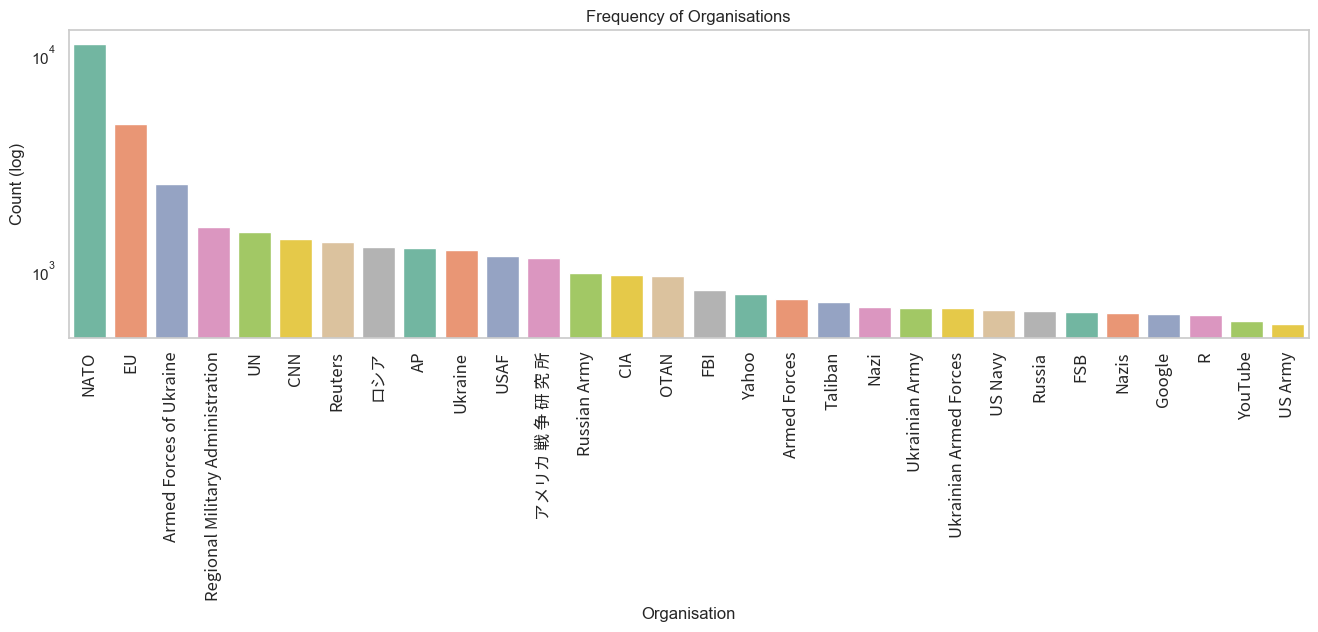}
    \caption{Most frequent named organization entities.}
    \label{fig:ner_organizations}
\end{figure}

Finally, about 11\% of the texts contain mentions of organization names, shown in figure \ref{fig:ner_organizations}. NATO is the most frequently discussed one, but surprisingly, the European Union appears before local institutions. News outlets and online platforms are also mentioned often (``CNN'', ``Reuters'', ``AP'', ``Yahoo'', ``Google'', ``YouTube''), as are U.S. military branches and secret services (``USAF'', ``CIA'', ``FBI'', ``US Navy'').

\section{Community detection}
Detecting communities of users associated with OSINT can be extremely helpful in detecting misinformation and determining trustworthiness, analyzing political stances, and generally gaining information about the spread of information. 

To implement this, we analyze users' mentions of each other (indicated by \texttt{@username}), or retweets of each others' messages. This results in a directed graph where edges are weighted based on the number of interactions/retweets. We find high indegrees (i.e. frequent mentions/retweets) for well-known OSINT accounts according to \cite{Hayman2023OpenSource} (e.g. \texttt{@ralee85, @oryxspioenkop, @defmon}). Then, we employ the Leiden algorithm to detect clusters within the graph. 

This results in a fairly high number of total clusters (95 when using mentions, 68 for retweets), but only six each that contain a salient number of members. We analyze these clusters further in terms of topics (modeled via Latent Dirichlet Allocation), partisanship, and misinformation, and find significant correlations for these properties. In effect, this means we can automatically determine OSINT communities that are focused on certain topics, and that display pro-Russian or pro-Ukrainian stances. Moreover, we confirm that certain communities are more likely to spread misinformation than others \cite{rode-hasinger-etal-2022-true}, giving us an important tool for misinformation detection. 

As a second type of analysis, we create a more sophisticated graph using the tweet and user features from the previous analyses as inputs. The edges between user/tweet nodes are then weighted with interaction scores, and node embeddings are learned using GraphSAGE \cite{hamilton2018inductiverepresentationlearninglarge}. Finally, the embedded nodes are clustered via Spectral clustering \cite{von2007tutorial}. Again, we find significant cluster correlations with partisanship and misinformation, but also engagement and offensive language, providing us with even deeper insights into user and network behavior. However, the distinction between high- and low-misinformation communities is not as clearcut here as in the previous analysis.

The full analysis is available under [REMOVED FOR REVIEW].


\section{Conclusion and future work}

In this paper, we conduct a range of analyses on an OSINT dataset about the Russo-Ukrainian war to obtain deeper insights into communication about this topic. We find correlations of sentiment and named entity distribution with war events. Named Entity Recognition also provides insights about actors and locations deemed relevant to the conflict. Partisanship detection reveals that while the majority of tweets are neutral, pro-Russian stances are surprisingly slightly more frequent than pro-Ukrainian ones. A preliminary misinformation detection confirms that most messages are probably trustworthy, but around 9\% are not; those are not spread evenly across partisanship. Finally, we analyze user networks with two different strategies and are able to detect sub-communities separated by partisanship, discussed topics, and amount of misinformation spread.

Several avenues remain open for deeper exploration:
\begin{itemize}
    \item Misinformation Dissemination: Further research is needed to explore the dynamics of how misinformation spreads through these networks, including temporal and multimodal factors that influence misinformation dissemination.
    \item Multimodal Content Analysis: The dataset includes a significant amount of image and video content that merits detailed examination to enhance our understanding of how visual OSINT is used and discussed.
    \item Geolocation Analysis: Investigating the use of geolocations in OSINT tweets could provide insights into how digital information correlates with real-world events and locations.
    \item Influential Actors: Targeted research into individuals and groups responsible for disseminating OSINT and misinformation could help identify and mitigate the impact of malicious actors in social media landscapes.
    \item Threaded Conversations: Current data collection methods have limitations in capturing threaded discussions, which are crucial for understanding detailed and controversial OSINT debates. Future methodologies should aim to capture these threaded conversations by penetrating deeper into users' interactions and discussions.
\end{itemize}

These areas of future research could significantly advance our understanding of the complexities of social media use in conflict contexts and improve strategies for combating misinformation.

\section*{Declaration on Generative AI}
 During the preparation of this work, the author(s) used GPT-4 in order to: Paraphrase and reword, improve writing style, abstract drafting. After using this tool, the authors reviewed and edited the content as needed and take full responsibility for the publication’s content. 

\small \bibliography{paper}



\end{document}